\documentclass[journal]{IEEEtran}
\usepackage{amsmath,amssymb,amsfonts}
\usepackage{graphicx}

    \usepackage[left = 1.95 cm, right = 1.95 cm, top = 1.95 cm, bottom = 1.95 cm]{geometry}

\usepackage{enumitem}    

\usepackage[english]{babel}
\usepackage[utf8]{inputenc}
 \usepackage[ruled,vlined]{algorithm2e}
 
 \usepackage{cite}

\ifCLASSINFOpdf
\else
\fi
\usepackage{amsmath,amssymb,amsfonts}

\usepackage{graphicx}
\graphicspath{%
    {converted_graphics/}
    {/}
}
  \newcommand{\MeijerG}[7]{G \begin{smallmatrix} #1 & #2 \\ #3 & #4 \end{smallmatrix} \left( \begin{smallmatrix} #5 \\ #6 \end{smallmatrix} \middle\vert #7 \right) }

  \makeatletter
\newcommand*{\rom}[1]{\expandafter\@slowromancap\romannumeral #1@}
\makeatother

\usepackage{xcolor}
  
\begin{document}
 \bstctlcite{IEEEexample:BSTcontrol}

\title{Maximizing Reliability  in Overlay Radio Networks with Time Switching and Power Splitting Energy Harvesting\\
}

 \author{Deemah H. Tashman,~\IEEEmembership{Member, IEEE}, Soumaya Cherkaoui,~\IEEEmembership{Senior Member, IEEE}, and Walaa Hamouda,~\IEEEmembership{Senior Member, IEEE}
\thanks{D. Tashman and S. Cherkaoui are with the Department of Computer  and Software Engineering, Polytechnique Montreal, Montreal, QC, Canada,  H3T 1J4 (e-mail: \{deemah.tashman, soumaya.cherkaoui\}@polymtl.ca).} 
\thanks{ W. Hamouda is with the Department of Electrical and Computer Engineering, Concordia University, Montreal,  QC, Canada, H3G 1M8, (e-mail:  hamouda@ece.concordia.ca).  }}

 \maketitle

\begin{abstract}
Cognitive radio networks (CRNs) are acknowledged for their ability to tackle the issue of spectrum under-utilization.  
In the realm of CRNs,  this paper investigates the energy efficiency issue and addresses the critical challenge of optimizing system reliability for overlay CRN access mode.  Randomly dispersed secondary users (SUs) serving as relays for primary users (PUs) are considered, in which one of these relays is designated to harvest energy through the time switching-energy harvesting (EH) protocol. Moreover, this relay amplifies-and-forwards (AF) the PU's messages and broadcasts them along with its own across cascaded $\kappa$-$\mu$ fading channels.  The power splitting protocol is another EH approach utilized by the SU and PU receivers to enhance the amount of energy in their storage devices. In addition, the SU transmitters and the SU receiver are deployed with multiple antennas for reception and apply the maximal ratio combining approach. The outage probability is utilized to assess both networks' reliability. Then, an energy efficiency evaluation is performed to determine the effectiveness of EH on the system. Finally, an optimization problem is provided with the goal of maximizing the data rate of the SUs by optimizing the time switching and the power allocation parameters of the SU relay.
\end{abstract}


\IEEEpeerreviewmaketitle

\section{Introduction}
\par\IEEEPARstart{C}{ognitive}  radio (CR) technology was developed to solve the problem of spectrum under-utilization induced by rigid spectrum allocation policies \cite{8910629,7955907,9533497,tashman2023security,10368012}. Demand for this technology is higher than ever before due to the expanding popularity of  5G, 6G,  and beyond wireless networks and their accompanying applications and services. Cognitive radio networks (CRNs) identify two types of networks; primary users (PUs) and secondary users (SUs) \cite{7275088}. PUs are users with permission to utilize a certain frequency range, whereas SUs are those without such permission. The SUs may use one of three access modes—interweave, underlay, or overlay—to access the licensed frequencies \cite{4840529}. SUs are permitted to use the licensed band in interweave mode as long as the bands are vacant \cite{9408651}. Using the underlay mode, PU and SU transmissions may occur simultaneously provided that SU interference is within a specified limit  \cite{9348134,7955106}. In the overlay method, the SUs gain access to the licensed band in exchange for serving as a relay node for the PUs. With the overlay access mode, the energy needed to relay the PUs' messages alongside the SUs' messages increases. Therefore, apart from the additional functions performed by SUs that need energy, the energy consumption challenge associated with the collaboration in overlay CRNs should be addressed. 

 The aforementioned challenge brings us to the role of energy harvesting (EH), which has lately emerged as a viable method for addressing the energy consumption issue, particularly for energy-constrained systems \cite{8269301,7501853,8187643,7914600,10368012}.  One of the effective EH techniques is simultaneous wireless information and power transfer (SWIPT) \cite{7955062}. This technique is based on the notion that radio frequency (RF) signals are composed of both energy and information \cite{8214104}.  In this case, the receiver can harvest energy from the same received messages.  However, to enable the SWIPT technique, the receiver must implement one of the following EH protocols: power splitting (PS), time switching (TS), or antenna selection (AS) \cite{10182973}. A TS-based communication mechanism typically divides its time frame into two time portions. The receiver gathers energy from the surrounding received signals during the first time slot. Afterward, it transmits the information along to its destination using the stored energy during the second time slot. Using a PS factor, the receiver in PS-EH divides the RF signal's energy into two segments: one is dedicated to EH, while the rest is utilized for processing the data  \cite{7317504,10278964}.  EH becomes increasingly critical in 5G networks and beyond, due to the enormous number of devices, especially vehicles \cite{9839316,triwinarko2021phy,9524882,9497103,mlika2021network}, which will require energy for communication. Furthermore, this expansion results in a heightened need for spectrum within the already under-utilized spectrum, a challenge that must be addressed.

To overcome the spectrum issue for vehicles, cognitive vehicular networks research has recently captured the public's attention since CRNs allow SUs operating as vehicles to access licensed frequency bands  \cite{9852474}. It is noteworthy to mention that previous research often assumed that the wireless links between vehicles followed standard channel behavior, such as the Rayleigh fading model. However, because of the mobility of devices, the modeling of signal propagation using these channels may not be accurate \cite{8820071}.  Therefore, cascaded fading channels have emerged as a reliable way of simulating signal propagation. Particularly, in scenarios where devices are in motion or when a large number of barriers stand between the transmitter and receiver \cite{9237455,9383093}. The received signal in cascaded channels is modeled as the product of a large number of rays reflecting off of objects  \cite{ghareeb2020statistical,9385753}.  Furthermore, it is widely acknowledged that cascaded channels are beneficial in simulating a wide range of systems, including multi-hop relaying systems and mobile-to-mobile/vehicle-to-vehicle (M2M/V2V) communication systems, to name a few \cite{9094381}.

Recent research has focused on improving the energy efficiency and security of the underlay CRN through the utilization of EH, such as \cite{9612017,9500621,9838746}, and the references therein. However, few studies have been conducted on employing EH for overlay CRNs.  For instance, PUs and SUs operate together in \cite{7342976}; the assisting SU derives power from the PUs' communications through the PS protocol. The probability of network outages and energy efficiency in both systems have been evaluated. Moreover, in \cite{8540871}, SUs undertake a TS-EH procedure, in which the outage probability and the overall system throughput have been evaluated. Furthermore, an overlay CRN was investigated in \cite{8436990}, in which the SU relays the PUs' messages in return for using the bands. In this work, the PU harvests energy from the SUs' transmissions to keep its own batteries charged. The PS factor has been optimized to boost SU and PU communication reliability. Recently, in \cite{9926102} the outage probability for an overlay CRN with the assistance of multiple EH relays has been investigated in terms of the outage probability over Rayleigh fading channels. The TS factor and the power allocation factor for the relay have been optimized to maximize users' throughput.

   To the authors' best knowledge, no prior study examined employing EH for an overlay CRN with multiple randomly distributed SUs considering cascaded fading channels.  Hence, this work investigates the overlay CRN reliability in the presence of multiple randomly distributed SUs functioning as amplify-and-forward (AF) and EH relays. These multiple SUs assist the PUs' communication over cascaded $\kappa$-$\mu$ fading channels. Moreover, the SUs are assumed to be randomly distributed according to a homogeneous Poisson point process (HPPP), in which one of the SUs is selected based on the Euclidean distance. The selected SU is assumed to have multiple antennas for reception and  harvests energy from the PUs' messages by adopting the TS protocol. Then, utilizing the gathered energy, this SU combines its own messages with the amplified PUs  messages and forwards them to the destinations.  Both destinations, i.e., the PU destination and the SU destination are assumed to harvest energy using the PS protocol to enhance their battery energy content. This energy is essential to compensate for all energy losses and to empower them for subsequent transmissions.  The reliability of the SUs and PUs networks is investigated in terms of the outage probability. Additionally,   the overall energy efficiency of the system is analyzed. In addition, we explore an optimization problem in which the TS and the power allocation parameters are optimized in order to maximize the rate of SUs while retaining the rate of PUs above a threshold.  It is noteworthy to mention that our system model and analyses are applicable to cognitive vehicular networks, which can be employed in the realm of smart transportation. This facilitates extensive communication among a large number of interconnected vehicles, thereby enabling connectivity and optimizing traffic management. Moreover, our motivation stems from the fact that energy consumption is a critical concern in vehicular networks, which is addressed in our work through EH to sustain power-hungry devices in urban environments.

The rest of the paper is organized as follows;  the system and channel models   are given in Section \rom{2}. In Section \rom{3}, the outage probability analysis  is presented.  Section \rom{4} includes the energy efficiency evaluation. The optimization problem   is presented in Section \rom{5}.  Section \rom{6} presents and discusses the numerical results. Finally,  conclusions are included in Section \rom{7}.

\begin{table}[htbp]\label{table1}
    \small
    \centering
    \caption{List of Main Symbols}
    \begin{tabular}{|c|c|p{4cm}|}
    \hline
    \textbf{Symbol} & \textbf{Definition}  \\
    \hline
    $L_R$ & Number of antennas at $R_k$  \\
    $L_S$ & Number of antennas at SU-Rx \\
    $j^{th}$ antenna & $R_k$ antenna index \\
    $l^{th}$ antenna & SU-Rx antenna index\\
    $P_T$ & PU-Tx transmission power \\
    $P_k$ & $R_k$ transmission power\\
    $\rho$ & TS factor \\
    $\eta$ & Energy conversion efficiency coefficient \\
    $\alpha$ & Path loss exponent \\
    $A_f$ & Power allocation factor \\
    $\nu_p$ & PS factor at PU-Rx \\
    $\nu_s$ & PS factor at SU-Rx\\  
    $n_p$ & $h_{RP}$ channel cascade level \\
    $n_s$ & $h_{RS}$ channel cascade level \\
    $\phi$ & Density of SUs \\
    $\kappa_{js}$ and $\mu_{js}$ &  $h_{RS}$ fading   parameters  \\
    $\kappa_{jp}$ and $\mu_{jp}$ &  $h_{RP}$ fading   parameters  \\
    $\lambda_p$ &   $h_{PR}$ fading   parameter \\
    $R_{thp}$ & PUs' transmission rate threshold \\
    $R_{ths}$ & SUs' transmission rate threshold \\
    $\mathcal{E}$ & Energy efficiency \\
    \hline  
    \end{tabular}
\end{table}

\IEEEpeerreviewmaketitle

 \section{System Model} 
 \label{system_model_section}
  \begin{figure}[t]
  \centering
  \includegraphics[width=0.8\linewidth]{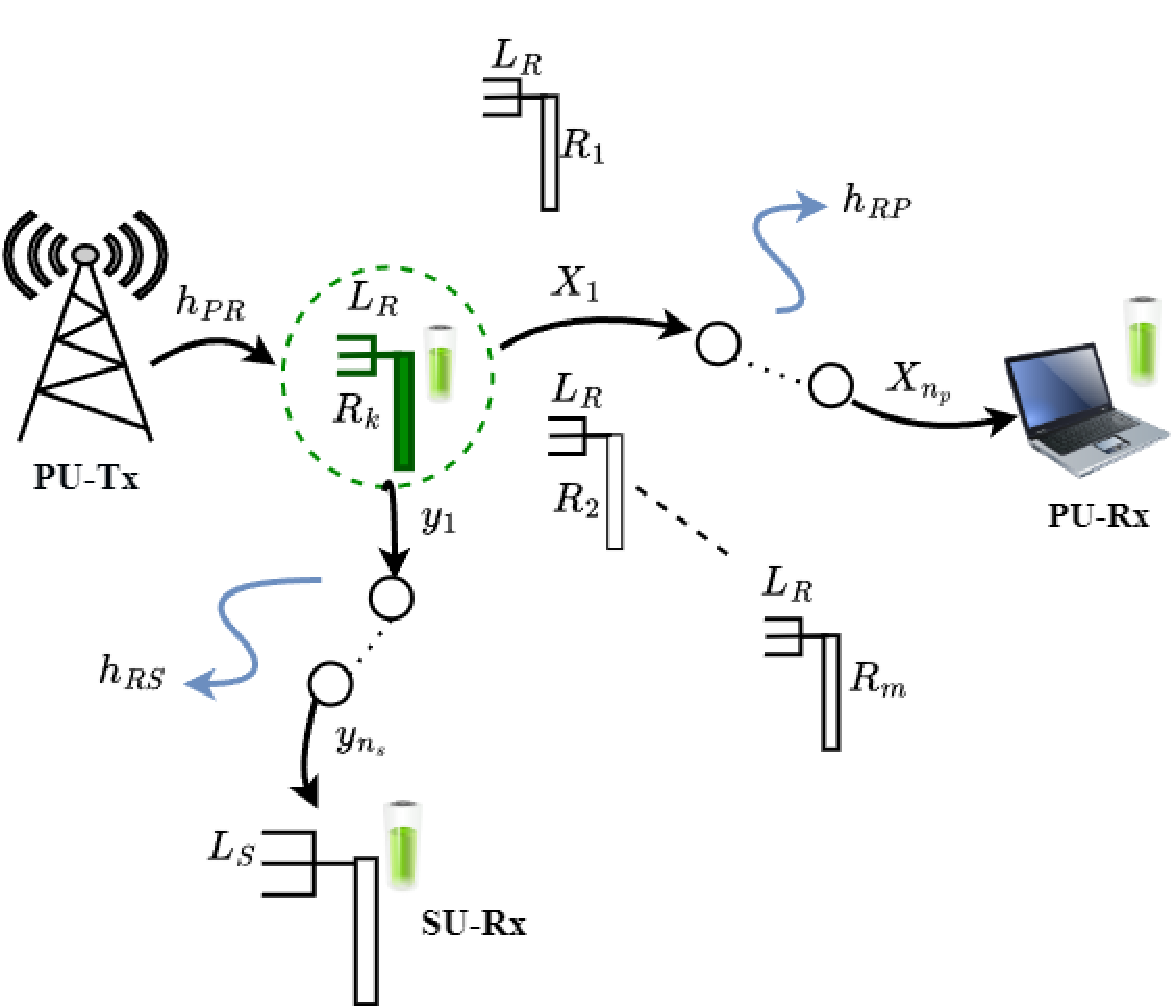}
  \caption{The system model.}
  \label{sys1}
\end{figure}
As shown in Figure \ref{sys1}, we assume two primary users (PUs), particularly, a PU transmitter (PU-Tx)  and a PU receiver (PU-Rx), are equipped with a single antenna and communicate with each other through the assistance of a number of relays. This is under the assumption that the PUs do not have a direct reliable channel of transmission and require the assistance of relays \cite{9839265,10093083}. These relays are secondary users (SUs) transmitters who have the intention of utilizing the licensed band in exchange for forwarding the PUs messages.   Moreover, these SUs  are distributed randomly according to a homogeneous Poisson point process (HPPP) \cite{9878110} with a density   $\phi$ \cite{tashman2021cascaded}.  In addition, one of these SUs is selected to receive the PU-Tx transmissions, from which it will harvest energy using the time switching (TS) protocol and store it in its storage device.  This selected SU functions as an  AF relay and uses the harvested energy to amplify the PUs messages. Furthermore, the relay combines its own messages with the PUs messages to eventually forward them to the PU-Rx and the SU receiver (SU-Rx). It is presumed that the SUs are distributed in an unbounded Euclidean space of dimension $U$.  It is worth mentioning that assuming multiple relays guarantees the continuity of the PUs and SUs transmissions. For instance, a distant  SU is selected when the closer SUs  suffer from unstable channels with the PU-Tx. Moreover, to improve the SUs' link reliability, we assume that the relay and the SU-Rx are  equipped with multiple antennas ($L_R$ and $L_S$, respectively) for reception. It is also assumed that the relay and the SU-Rx utilize the maximal ratio combining (MRC) technique to improve the received signal-to-noise ratio (SNR). Additionally, we assume that both receivers, i.e., PU-Rx and SU-Rx are energy-constrained devices that need to harvest energy from the relay's messages to be stored in their storage devices. This power can be utilized for compensating for any energy loss occurred and for empowering following transmissions \cite{9612017}. These receivers harvest energy using the power splitting (PS) protocol. Hence, two EH receiving schemes are considered in this  system model. In addition, it is assumed that the receiving connections ($h_{RP}$ and $h_{RS}$) are affected by a number of dispersed objects and scatters. As a result, it is more reasonable to hypothesize that they adhere to cascaded channels. On the contrary, it is assumed that the relay is in close proximity to PU-Tx and that the link connecting them ($h_{PR}$) encounters minimal obstacles; therefore, it is more likely to adhere to a single fading model \cite{9094381,9612017,9500621}. This allows us to evaluate the system model in the worst-case situation, in which the receiving connections suffer from severe fading. Moreover, this demonstrates a more practical scenario concerning cognitive vehicular networks \cite{9237455}. The main symbols  employed throughout the paper and their definition are summarized in Table \ref{table1}.

\begin{figure*}
 \centering
 \includegraphics[width=0.8\linewidth]{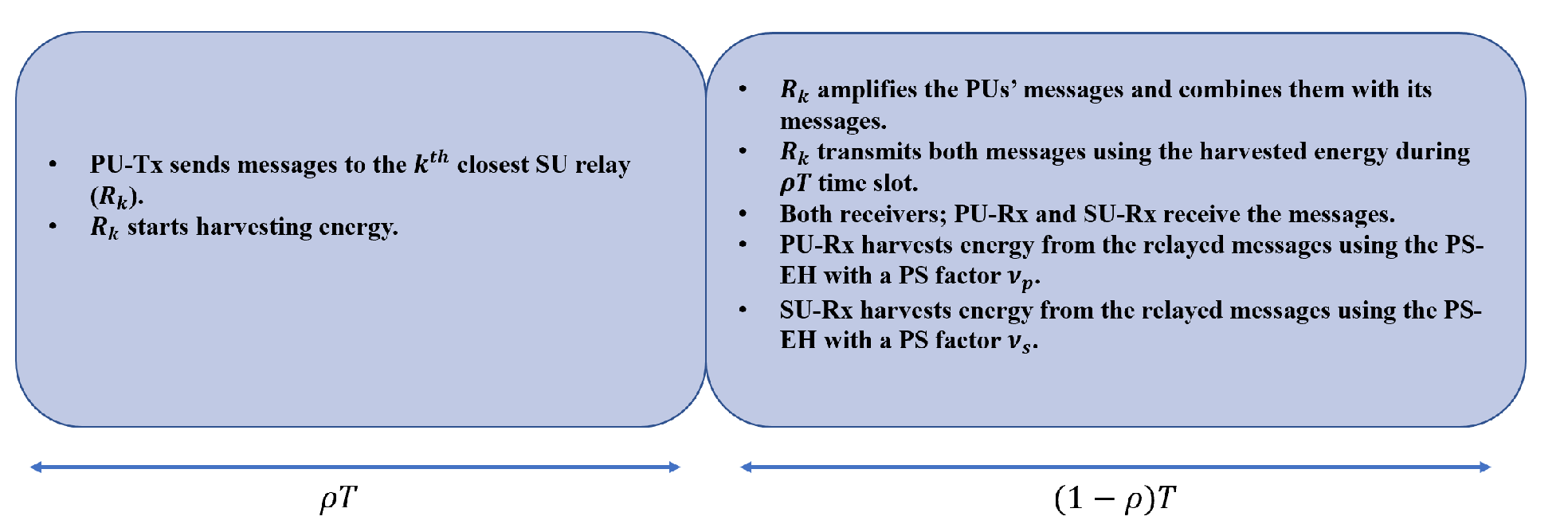}
 \caption{Frame structure illustration of the EH and messages' relay process; The time interval $T$ is partitioned into two time slots: $\rho T$ and $(1-\rho)T$. Messages' transmission by PU-Tx takes place during $\rho T$, whereas amplification and relaying of messages, as well as the EH by the receivers, occur during $(1-\rho) T$ slot.
}
 \label{sys2}
\end{figure*}

 Since the $k^{th}$ closest relay utilizes the TS technique for harvesting energy from PU-Tx, Figure \ref{sys2} shows the detailed process split according to the TS factor ($\rho$). For the first time portion, i.e., $\rho T$, the chosen relay  $(R_k)$ harvests energy from the PU-Tx messages. The harvested energy $(E_p)$ is given by 
  \begin{IEEEeqnarray}{lCr} \label{EHmain}
E_p= \frac{ \rho \eta P_T T \left|h_{PR}\right|^2}{d^{\alpha}},
\end{IEEEeqnarray}
\noindent where $P_T$ is PU-Tx transmission power,  $0<\rho<1$ is the TS  factor, $\eta$ represents the energy conversion efficiency coefficient, $d$ is the distance of a randomly distributed SU from PU-Tx, $\alpha$ is the path loss exponent,   and $T$ is the transmission time slot. Given  (\ref{EHmain}), the transmission power $(P_k)$ at the $k^{th}$ random SU $(R_k)$ is given by 
   \begin{IEEEeqnarray}{lCr}  \label{pr}
P_k= \frac{E_p}{(1-\rho)T}= \frac{\rho \eta P_T   \left|h_{PR}\right|^2}{(1-\rho) d^{\alpha}}.
\end{IEEEeqnarray}
\noindent The received message at the $j^{th}$ antenna of $R_k$  is given by
\begin{eqnarray}\label{msgrelay}
y_{Rk}^{(j)}=\sqrt{\frac{P_T}{d^{\alpha}}} h_{PR}^{(j)} x_p+n_R^{(j)},
\end{eqnarray}
\noindent where    $x_p$ is the PUs' transmitted message, $h_{PR}^{(j)}$ is the channel gain between PU-Tx and the $j^{th}$ antenna of $R_k$, and $n_R$ is the additive white Gaussian noise (AWGN) at  the $j^{th}$ antenna of $R_k$ with a zero mean and a variance $N_0$. Then, during the second time slot  $\left((1-\rho)T\right)$,   $R_k$  performs the amplification of the  PUs' messages and adds them to  its   messages to be forwarded   to PU-Rx and SU-Rx.  Hence, the received message at the PU-Rx   is given by
\begin{eqnarray} \label{msgpu}
y_P= \Lambda y_{Rk}^{(j)} h_{RP} +\sqrt{(1-A_f)(1-\nu_p)P_k} h_{RP} x_s+n_P,
\end{eqnarray}
\noindent where $x_s$ is the transmitted SUs' messages and $n_P$ is the AWGN at PU-Rx with a zero mean and a variance $N_0$.  $A_f$ represents the power allocation factor, in which $A_f P_k$ is allocated to forward the PUs messages, while   $(1-A_f)P_k$ is employed to transfer the SUs' messages. Moreover, $\nu_p$ is the PS factor at PU-Rx and   $\Lambda$ is the amplification factor of $R_k$ and it is given by
\begin{eqnarray} \label{beta}
\Lambda=\sqrt{\frac{A_f P_k}{\frac{P_T G_{PR}}{d^{\alpha}}+N_0}},
\end{eqnarray}
\noindent   where $G_{PR}= \sum_{j=1}^{L_R}  g_{PR}^{(j)}$ with $g_{PR}^{(j)}=\left|h_{PR}^{(j)}\right|^2$. The noise variance $N_{0}$ in (\ref{beta})  can be ignored compared to the term $\frac{P_T G_{PR}}{d^{\alpha}}$ at high SNR \cite{7342976}. Therefore, (\ref{beta}) is approximated as
 \begin{eqnarray} \label{betaapp}
\Lambda\approx\sqrt{\frac{A_f P_k}{\frac{P_T G_{PR}}{d^{\alpha}} }}.
\end{eqnarray}
\noindent Substituting (\ref{msgrelay}) into (\ref{msgpu}), the received message at the PU-Rx is expressed as
\begin{eqnarray}  \label{ydlast}
y_P&=& \Lambda \sqrt{\frac{P_T}{d^{\alpha}}} h_{PR}^{(j)} x_p h_{RP} + \sqrt{(1-A_f)P_k} h_{RP} x_s \nonumber \\ 
&&  + \Lambda  h_{RP} n_P  + n_R^{(j)}.
\end{eqnarray}
\noindent It is assumed that each of the receivers treat the other's messages as interference. Hence, given   (\ref{betaapp}) and (\ref{ydlast}), the instantaneous  received signal-to-interference-plus-noise ratio (SINR)   at PU-Rx is expressed as
 \begin{eqnarray}  \label{snrd}
\gamma_P &=& \frac{(1-\nu_p)\Lambda^2 g_{RP}\frac{P_T}{d^{\alpha}}G_{PR}   }{N_0 g_{RP} \Lambda^2+g_{RP} (1-A_f) P_k+N_0} .
\end{eqnarray}
\noindent where $g_{RP}=\left|h_{RP}\right|^2$. Substituting   (\ref{pr}) and (\ref{betaapp})  into (\ref{snrd}) and performing mathematical manipulations yields
 \begin{eqnarray}  \label{snrd2}
\gamma_P &=& \frac{a \frac{g_{RP} G_{PR}}{d^{\alpha}}}{b g_{RP}  + c \frac{g_{RP} G_{PR}}{d^{\alpha}} +N_0} ,
\end{eqnarray}
\noindent where $a=\frac{(1-\nu_p)A_f\rho \eta P_T}{1-\rho}$, $b=\frac{N_0A_f\rho \eta }{1-\rho}$, and $c=\frac{(1-A_f) \rho \eta P_T}{1-\rho}$. Moreover, the received message at the $l^{th}$ antenna of   SU-Rx  is given by
\begin{eqnarray} \label{ysfirst} 
y_S^{(l)}&=   \sqrt{(1-A_f)(1-\nu_s)P_k} h_{RS}^{(l)} x_s+n_s^{(l)}+\Lambda y_{R_{k}}^{(j)} h_{RS}^{(l)} , \nonumber \\
\end{eqnarray}
\noindent where $\nu_s$ is the PS factor at SU-Rx, $h_{RS}^{(l)}$ is the channel gain between $R_k$ and the $l^{th}$ antenna of SU-Rx,  and  $n_s^{(l)}$ is the AWGN at the $l^{th}$ antenna of SU-Rx with a zero mean and a variance $N_0$. Using (\ref{ysfirst}), the received SINR at SU-Rx is given as
 \begin{eqnarray}  \label{snrs}
\gamma_S &=& \frac{q \frac{G_{RS} G_{PR}}{d^{\alpha}}}{e G_{RS}  + w \frac{G_{RS} G_{PR}}{d^{\alpha}} +N_0} ,
\end{eqnarray}
\noindent where $q=\frac{(1-\nu_s)(1-A_f)\rho \eta P_T}{1-\rho}$, $G_{RS}= \sum_{l=1}^{L_S}  g_{RS}^{(l)}$ with $g_{RS}^{(l)}=\left|h_{RS}^{(l)}\right|^2$,  $e=\frac{N_0A_f\rho \eta }{1-\rho}$, and $w=\frac{A_f \rho \eta P_T}{1-\rho}$. Accordingly,  the data rate achieved at PU-Rx and SU-Rx are given, respectively, as 
 \begin{eqnarray} \label{rate}
 R_p=(1-\rho)T \log_2 (1+\gamma_P), 
 \end{eqnarray}
 \begin{eqnarray} \label{ratesu}
 R_s=(1-\rho)T \log_2 (1+\gamma_S).
 \end{eqnarray}
 
We assume that the links $h_{RP}$ and $h_{RS}$ follow the cascaded $\kappa$-$\mu$ fading channels. Hence, $h_{RP}=\prod_{i=1}^{n_p} X_i$, where $X_i$ follows the $\kappa$-$\mu$ fading model and $n_p$ is the cascade level of the channel $h_{RP}$. Therefore, the probability density function (PDF)   of $g_{RP}$ is given by \cite{9094381}
\begin{IEEEeqnarray}{lcr}
\label{pdf_main-nnn-fgrp}
f_{g_{RP}} (x)=\sum_{r_1=0}^{\infty} \sum_{r_2=0}^{\infty} \cdots \sum_{r_{n_p}=0}^{\infty} \frac{c_{xp}}{2}  x^{\mu_{1p} +r_1-1}    \nonumber \\  
 \times    \MeijerG{n_p}{0}{0}{n_p}{-}{1-\beta_P}{{  }{x \prod_{j=1}^{n_p}   \mu_{jp} \left(1+\kappa_{jp}\right)  }}    ,  
\end{IEEEeqnarray} 
\noindent where $\beta_p = \mu_{1p}-\mu_{2p} +r_1-r_2+1,\cdots,\mu_{1p} -\mu_{n_p} +r_1-r_{n_p}+1,1$  and
\begin{align*}
c_{xp}=& 2 \prod_{j=1}^{n_p} \left[  \frac{  \left[2\mu_{jp}  \sqrt{\kappa_{jp}\left(1+\kappa_{jp}\right)} \right]^{2r_{j}+\mu_{jp} -1}  }{(r_{j})! 2^{2r_{j}+\mu_{jp}  -1}\kappa_{jp}^{\frac{\mu_{jp}  -1}{2}}  }  \right]  
\nonumber \\  
\;\;\;\;\;\;\;\;\; &\times  
\prod_{j=1}^{n_p} \left[\frac{\left[\mu_{jp} \left(1+\kappa_{jp} \right)\right]^{\mu_{1p} -\mu_{jp} +r_1-r_{j}} }{\exp\left(\kappa_{jp}\mu_{jp} \right) \Gamma\left(r_{j}+\mu_{jp}\right)}\right]
\nonumber \\  
\;\;\;\;\;\;\;\;\; &\times  
\prod_{j=1}^{n_p} \left[ \mu_{jp} \left(1+\kappa_{jp}\right)^{\frac{\mu_{jp} +1}{2}}\right],
\end{align*}
\noindent with $\kappa_{jp}$ and $\mu_{jp}$ are the   fading channel parameters for $j=1,2,\cdots,n_p$. 
Similarly, the channel gain $h_{RS}^{(l)}$ follows the cascaded $\kappa$-$\mu$ distribution. Hence, $h_{RS}^{(l)}=\prod_{i=1}^{n_s} y_{i}^{(l)}$, with $n_s$ being the channel cascade level and $y_{i}^{(l)}$ is a random variable following the $\kappa$-$\mu$ fading model. The PDF of $G_{RS}$ with $\kappa_{js}$ and $\mu_{js}$ being the fading parameters for $j=1,2,\cdots,n_s$ is given as  \cite{9385753}
\begin{IEEEeqnarray}{lcr}
\label{pdf_main-nnn}
f_{G_{RS}} (x)=\sum_{v_1=0}^{\infty} \sum_{v_2=0}^{\infty} \cdots \sum_{v_{n_s}=0}^{\infty} \frac{c_{xs}}{2 L_S^{\mu_{1s} L_{S}+v_1}}   \nonumber \\  
 \times    \MeijerG{0}{n_s}{n_s}{0}{\beta_s}{-}{\frac{  L_S}{x \prod_{j=1}^{n_s}   \mu_{js} L_{jS}\left(1+\kappa_{js}\right)  }}    x^{\mu_{1s} L_S+v_1-1}   ,   \nonumber \\
\end{IEEEeqnarray} 
 \noindent where $\beta_s = \mu_{1s} L_S-\mu_{2s} L_S+v_1-v_2+1,\cdots,\mu_{1s} L_S-\mu_{n_s} L_S+v_1-v_{n_s}+1,1$  and
\begin{align*}
c_{xs}=& 2 \prod_{j=1}^{n_s} \left[  \frac{  \left[2\mu_{js} L_{jS} \sqrt{\kappa_{js}\left(1+\kappa_{js}\right)} \right]^{2v_j+\mu_{js} L_{jS}-1}  }{(v_j)! 2^{2v_j+\mu_{js} L_{jS} -1}\kappa_{js}^{\frac{\mu_{js} L_{jS} -1}{2}}  }  \right]  
\nonumber \\  
\;\;\;\;\;\;\;\;\; &\times  
\prod_{j=1}^{n_s} \left[\frac{\left[\mu_{js} L_{jS}\left(1+\kappa_{js} \right)\right]^{\mu_{1s} L_S-\mu_{js} L_{jS}+v_1-v_j} }{\exp\left(\kappa_{js}\mu_{js} L_{jS}\right) \Gamma\left(\nu_j+\mu_{js} L_{jS}\right)}\right]
\nonumber \\  
\;\;\;\;\;\;\;\;\; &\times  
\prod_{j=1}^{n_s} \left[ \mu_{js} L_{jS} \left(1+\kappa_{js}\right)^{\frac{\mu_{js} L_{jS} +1}{2}}\right]
 \cdot
\end{align*}
\noindent In addition, it is assumed that the link between the PU-Tx and the relay $R_k$ $(h_{PR})$ follows the single  Rayleigh model, which is a special case of the $\kappa$-$\mu$ model. Therefore, the PDF and the cumulative distribution function (CDF) for the power channel gain of $h_{PR}$, i.e., $G_{PR}$ follow the exponential distribution with $\lambda_p$ being the fading coefficient as
\begin{eqnarray} \label{hPRpower}
f_{G_{PR}} (x)&=&  \frac{\lambda_p^{L_R}}{(L_R-1)!} x^{L_R-1} \exp(-\lambda_p x),
\end{eqnarray} 
\begin{eqnarray} \label{hPRpower}
F_{G_{PR}} (x)&=& 1-\sum_{i=0}^{L_R-1} \frac{\lambda_p^{i} x^i\exp(-\lambda_p x)}{i!}.
\end{eqnarray} 
\noindent  As mentioned in the system model description in section \ref{system_model_section}, the $k^{th}$   nearest SU   to    PU-Tx  will be selected to forward the messages. This is performed by measuring the Euclidean distance from PU-Tx to each of the SUs.  The PDF of the path loss $d^{\alpha}$ for the $k^{th}$ nearest SU is  distributed  as \cite{1512427}
  \begin{eqnarray} \label{pdf_pl}
f_{d^{\alpha}} (x) &=& \exp\left({-A_e x^{\delta}}\right) \frac{\delta A_e^{k} x^{\delta k -1}}{\Gamma(k)}, 
\end{eqnarray} 
 \noindent where $A_e=\pi \phi$ and $\delta=\frac{U}{\alpha}$. 
 
\section{outage probability} \label{section_3}
The networks' reliability is investigated in terms of the outage probability $(OP)$, which is defined as the probability that the data rate is below a predetermined rate level. In our system model, evaluating $OP$ is critical since each network treats the other network's messages as interference. The outage probability is expressed as
\begin{eqnarray} \label{op1}
OP= \mathbb{P} (R_i \leq R_{thi}),
\end{eqnarray}
\noindent for $i=\{p,s \}$. $R_{thp}$ represents the rate threshold for the PUs' transmissions, while $R_{ths}$ is the threshold for the SUs' communication.  $\mathbb{P}(\cdot)$ represents the probability operator.  

\subsection{ Outage Probability of the Primary Users' Network}
The outage probability for the PUs' link $(OP_P)$ is evaluated in this section.  Using  (\ref{snrd}), (\ref{rate}), and (\ref{op1}), $OP_P$ is given by  
\begin{IEEEeqnarray} {lcr}
 \label{op_main}
OP_P=  \mathbb{P}\left(a g_{RP} g_{PR} \leq bJ g_{RP} d^{\alpha} + Jc g_{RP} g_{PR} +N_0 J d^{\alpha} \right)\nonumber \\
= \mathbb{P} \left(g_{PR} \leq \frac{bJ }{a-Jc}d^{\alpha}+\frac{N_0J }{(a-Jc)} L    \right)\nonumber \\
=\int_{0}^{\infty}  \int_{0}^{\infty} F_{g_{PR}}\left(c_1 z+c_2 l\right) f_{d^{\alpha}}(z) f_L(l)dz dl,              
 \end{IEEEeqnarray}
\noindent where $J=2^{\frac{R_{thp}}{(1-\rho)T}}-1$, $c_1=\frac{bJ}{a-Jc}$, $c_2=\frac{N_0 J}{a-Jc}$,  and $L=\frac{d^{\alpha}}{g_{RP}}$. The PDF of  $L$ should be first evaluated  as
\begin{eqnarray} \label{pdfL}
f_{L}(x)&=&\int_0^{\infty} y f_{d^{\alpha}} (xy) f_{g_{RP}} (y) dy \cdot
\end{eqnarray}
\noindent Substituting  (\ref{pdf_main-nnn-fgrp}) and (\ref{pdf_pl}) into (\ref{pdfL})   yields
\begin{IEEEeqnarray} {lcr}
  \label{pdfL2}
f_{L }(x)= \sum_{r_1=0}^{\infty} \sum_{r_2=0}^{\infty} \cdots \sum_{r_{n_p}=0}^{\infty}  \frac{\delta A_e^kc_{xp}}{2\Gamma(k)} x^{\delta k-1}\int_0^{\infty} y^{\delta k+\mu_{1p}+r_1-1} \nonumber \\   \times \MeijerG{1}{0}{0}{1}{-}{0}{A_e x^\delta y^\delta} \MeijerG{n_p}{0}{0}{n_p}{-}{1-\beta_p}{y\prod_{j=1}^{n_p}\mu_{jp}(1+\kappa_{jp}) }  dy. \nonumber \\
\end{IEEEeqnarray}
\noindent Using \cite [eq. (2.24.1.1)]{prudnikov1990integrals}, the PDF of $L$ is found as
\begin{IEEEeqnarray} {lcr}
f_{L }(x) = \sum_{r_1=0}^{\infty} \sum_{r_2=0}^{\infty} \cdots \sum_{r_{n_p}=0}^{\infty}  \xi x^{\delta k-1}        \nonumber \\  \times  \MeijerG{1}{\delta n_p}{\delta n_p}{1}{\frac{-\delta k-\mu_{1p}-r_1+\beta_p}{\delta}}{0}{\frac{A_e x^\delta \delta^{\delta n_p}}{\left(\prod_{j=1}^{n_p} \mu_{jp}(1+\kappa_{jp})\right)^\delta  }  }, \nonumber \\
\end{IEEEeqnarray}
\noindent where $\xi=\frac{A_e^kc_{xp}\delta^{n_p(\delta k+\mu_{1p}+r_1+1)+0.5} }{2\Gamma(k) (2\pi)^{(\delta-1)0.5}} \\ \times \left(\prod_{j=1}^{n_p} \mu_{jp}(1+\kappa_{jp}) \right)^{-\delta k-\mu_{1p}-r_1}$. The  outage probability of the PUs' link given  in (\ref{op_main})  is expressed as
\begin{IEEEeqnarray} {lcr}\label{opfinalpu}
\small
OP_P= \sum_{r_1=0}^{\infty} \sum_{r_2=0}^{\infty} \cdots \sum_{r_{n_p}=0}^{\infty} \frac{\xi \delta A_e^{k}}{\Gamma(k)}\int_{0}^{\infty}  \int_{0}^{\infty} \exp\left({-A_e z^{\delta}}\right) { z^{\delta k -1}}{}    \nonumber \\ \times  l^{\delta k-1} \left[1-\sum_{i=0}^{L_R-1} \frac{\lambda_p^{i} (c_1z+c_2l)^i\exp\left(-\lambda_p (c_1z+c_2l)\right)}{i!}\right]  \nonumber \\ \times     \MeijerG{1}{\delta n_p}{\delta n_p}{1}{\frac{-\delta k-\mu_{1p}-r_1+\beta_p}{\delta}}{0}{\frac{A_e  \delta^{\delta n_p}l^\delta}{\left(\prod_{j=1}^{n_p} \mu_{jp}(1+\kappa_{jp})\right)^\delta }  } dz dl  \nonumber \\ =1- \sum_{r_1=0}^{\infty} \sum_{r_2=0}^{\infty} \cdots \sum_{r_{n_p}=0}^{\infty}\frac{\xi\delta A_e^{k}}{\Gamma(k)}   S_1 S_2 ,
\end{IEEEeqnarray}
\noindent where $S_1$ and $S_2$ are respectively given by
\small
\begin{IEEEeqnarray}{lcr}
S_1=\sum_{i=0}^{L_R-1}c_1^i \sum_{j=0}^{i} \binom{i}{j} \left(\frac{c_2 }{c_1}\right)^{i-j} \int_{z=0}^{\infty} z^{j+\delta k-1}   \nonumber \\ \times \MeijerG{1}{0}{0}{1}{-}{0}{\lambda_{p}c_1z  }  \MeijerG{1}{0}{0}{1}{-}{0}{A_e z^\delta  }   dz,
\end{IEEEeqnarray}
\normalsize
\begin{IEEEeqnarray}{lcr}
\small
S_2=\int_{l=0}^{\infty} l^{i-j+\delta k-1}   \MeijerG{1}{0}{0}{1}{-}{0}{\lambda_{p}c_2 l  }  \\ \nonumber   \times \MeijerG{1}{\delta n_p}{\delta n_p}{1}{\frac{-\delta k-\mu_{1p}-r_1+\beta_p}{\delta}}{0}{ \frac{A_e  \delta^{\delta n_p}l^\delta}{\left(\prod_{j=1}^{n_p} \mu_{jp}(1+\kappa_{jp})\right)^\delta }   }   dl.
\end{IEEEeqnarray}
\normalsize
\noindent Using \cite [eq.(2.24.1.1)]{prudnikov1990integrals}, $S_1$  and $S_2$  are respectively solved  as
\small
\begin{IEEEeqnarray}{lcr} \label{I1foras}
S_1=\sum_{i=0}^{L_R-1}c_1^i \sum_{j=0}^{i} \binom{i}{j} \left(\frac{c_2 }{c_1}\right)^{i-j} \frac{\delta^{j+\delta k-0.5}}{(2\pi)^{(\delta-1)0.5} \left(c_1 \lambda_{p}\right)^{j+\delta k}} \nonumber \\    \times \MeijerG{1}{\delta}{\delta}{1}{D_{i}(\delta,1-j-\delta k)}{0}{ \frac{A_e \delta^\delta}{(\lambda_{p}c_1)^{\delta} }   },
\end{IEEEeqnarray}
\begin{IEEEeqnarray}{lcr} \label{I2foras} 
\small
S_2=\frac{\delta^{i-j+\delta k-0.5}}{(2\pi)^{(\delta-1)0.5} \left(c_2 \lambda_{p}\right)^{i-j+\delta k}}  \\ \nonumber  \times \MeijerG{1}{\delta n_p+\delta}{\delta n_p+\delta}{1}{\eta_g}{0}{ \frac{{A_e  \delta^{\delta+\delta n_p}l^\delta}{ }}{\left(\lambda_p c_2 \left(\prod_{j=1}^{n_p} \mu_{jp}(1+\kappa_{jp})\right)^\delta\right)^\delta}  },
\end{IEEEeqnarray}
\normalsize
\noindent where $\eta_g=\frac{-\delta k-\mu_{1p}-r_1+\beta_p}{\delta}, \frac{1-i+j-\delta k}{\delta}$ and   $D_i(\delta,1-j-\delta k)=\frac{1-j-\delta k}{\delta}, \frac{2-j-\delta k}{\delta},\cdots,\frac{\delta-j-\delta k}{\delta}$.
\normalsize
\subsection{Outage Probability of the Secondary Users' Communication}
As stated at the beginning of section \ref{section_3}, the SU-Rx considers the PUs' messages as interference that is degrading their transmissions, and thus it is essential to study the outage probability of their link. The outage probability of the SUs link $(OP_S)$ is evaluated using  (\ref{snrs}),   (\ref{ratesu}), and  (\ref{op1})  as
\begin{eqnarray} \label{op_main_su}
OP_S&=& \mathbb{P} \left(g_{PR} \leq d_1 {\frac{d^{\alpha}}{G_{RS}} }+d_2   {d^{\alpha}}{}     \right)                ,
\end{eqnarray}
 \noindent where $d_1=\frac{\epsilon_e N_o}{q-\epsilon_e w}$, $d_2=\frac{\epsilon_e e}{q-\epsilon_e w}$, and $\epsilon_e=2^{\frac{R_{ths}}{(1-\rho)T}}-1$.
\noindent Following the same procedure employed  to find (\ref{opfinalpu}), the $OP$ of the SUs' link is expressed as
\begin{IEEEeqnarray} {lcr}
OP_S =1-\sum_{i=0}^{L_R-1} \sum_{v_1=0}^{\infty} \sum_{v_2=0}^{\infty} \cdots \sum_{v_{n_s}=0}^{\infty}  \frac{d_3\delta A_e^{k} \lambda_p^i}{\Gamma(k) i!}I_1 I_2 , \nonumber \\
\end{IEEEeqnarray}
\noindent where $d_3=\frac{c_{xs}\delta^{1.5+\delta k} A_e^k \prod_{j=1}^{L_S}L_{jS}\mu_{js}(1+\kappa_{js})}{2L_S^{-\delta k}\Gamma(k) (2\pi)^{(\delta-1)0.5}}$. Similar to finding $S_1$ and $S_2$, $I_1$ and $I_2$ are expressed, respectively as
\begin{IEEEeqnarray}{lcr}
\small
I_1=\frac{\delta^{\delta k-0.5}}{(2\pi)^{(\delta-1)0.5} \left(d_2 \lambda_{p}\right)^{\delta k}}  \\ \nonumber  \times \MeijerG{1}{\delta}{\delta}{1}{\Delta(\delta,1-\delta k)}{0}{ \frac{A_e \delta^\delta}{(\lambda_{p}d_2)^{\delta} }   }, 
\end{IEEEeqnarray}
\small
\begin{IEEEeqnarray}{lcr}
 I_2=\frac{\delta^{\delta k-0.5}}{(2\pi)^{(\delta-1)0.5} \left(d_1 \lambda_{p}\right)^{i-j+\delta k}} \\ \nonumber   \times \MeijerG{1}{\delta n_s+\delta}{\delta n_s+\delta}{1}{\eta^\prime}{0}{ \frac{A_e L_S^\delta \delta^{\delta n_s+\delta}}{\left(\prod_{j=1}^{n_s}L_{jS} \mu_{js}(1+\kappa_{js})\lambda_{p}  d_1\right)^{\delta} }   },   
\end{IEEEeqnarray}
\normalsize
\noindent where $\eta^\prime=\frac{-\delta k-\mu_{1s}L_S-v_1+\beta_s}{\delta},\frac{1-i+j-\delta k}{\delta}$ and $\Delta(\delta,1-\delta k)=\frac{1-\delta k}{\delta}, \frac{2-\delta k} {\delta},\cdots,\frac{\delta-\delta k}{\delta}$.
 
\section{Energy Efficiency}  
To progress towards an EH-based green communication system, one must evaluate the energy efficiency of the considered system. Before we can analyze the energy efficiency, we must determine the total throughput. The throughput is defined as the sum of the average target transmission rates for PUs and SUs that are successfully attained\cite{8540871}, \cite{liu2016two}. Consequently, the system throughput $(\tau)$ for this overlay CRN is represented as 
 \begin{eqnarray}
 \tau=\tau_P + \tau_S,
 \end{eqnarray}
\noindent where $\tau_P$ is the throughput of the PUs' system and $\tau_S$ is the throughput of the SUs'  system, and they are given, respectively as
\begin{eqnarray}
\tau_P=(1-OP_P) R_{thp} (1-\rho), \nonumber \\
\tau_S=(1-OP_S) R_{ths} (1-\rho).
\end{eqnarray}

The energy efficiency $(\mathcal{E})$ is defined as  the ratio of the total amount of the delivered data to the total amount of the energy consumed as
\begin{eqnarray}
\mathcal{E}=\frac{\tau}{\left(\rho +\nu_p +\nu_s\right) P_T}.
\end{eqnarray}
\noindent Several parameters can be used to evaluate the system energy efficiency through this expression. Further explanations are provided in the numerical results section.

 \section{Optimization Problem} 
The time switching factor  $(\rho)$ and the power allocation factor of the relay $(A_f)$ both should be optimized to maximize the users' data rate. Optimizing $\rho$ assists in  managing the time slots dedicated to energy harvesting and amplifying the PUs messages and forwarding both users' transmissions.  Additionally, optimizing $A_f$ controls the portion of the power that the relay has to reserve for forwarding the PUs' messages and for transmitting the SUs' messages. Since each network considers the other network's messages as interference, optimizing $A_f$ would also control the amount of interference imposed on each receiver. 
 
 In this section, we optimize $\rho$ and $A_f$ with the objective of maximizing the SUs' rate (or reducing the $OP_S$), while ensuring that the PUs' data rate is maintained above an acceptable threshold $(R_{pt})$. This constraint ensures that regardless of  the amount of maximization achieved  for the SUs' rate, the PUs' reliability is maintained above the required level.  Given this, the optimization problem is formulated as
 \begin{align} \label{opti-prob}
   \mathcal{P}1: \;\;  & \underset{ \rho, A_f}{\text{max}}
    & & R_s \\
    & \text{s.t.} 
    & &  \label{firstrho} 0<\rho<1, &  \\
      & &&  \label{firstalpha} 0<A_f<1, &  \\
    & &&  \label{firstrate} R_p \geq R_{pt}. &  
  \end{align} 
 \noindent It is noticed that this is a non-linear mixed-integer optimization problem, and thus it is non-convex. However, it is a biconvex optimization problem  in  $\rho$  and $A_f$.  Particularly, setting $\rho$ to a constant value, the problem becomes convex in $A_f$, and similarly, for a fixed value of $A_f$, it becomes  convex of $\rho$.    This  can be proven by several methods, such as plotting the functions on Matlab. Similar to \cite{7342976} and \cite{9926102}, such a  problem   can be solved using the approach described in Algorithm 1, in which  $G$ represents the biconvex set of $A_f$ and  $\rho$, and $F$ represents the objective function in (\ref{opti-prob}).  
\begin{algorithm}
\caption{Algorithm for solving a biconvex optimization problem}
\KwIn{Initial parameters ${A_f}_0$, $\rho_0$}

\KwOut{Maximum $R_s$ and optimized   ${A_f}^\ast$ and $\rho^\ast$}
\BlankLine
Initialize $A_f = {A_f}_0$, $\rho = \rho_0$, and $i=0$\;

\Repeat{convergence}{
    Fix $\rho$ at $\rho_0$\;
    
    Calculate the optimal value of $A_f$ (${A_f}^\ast$) for the convex problem by $\max\{F(A_f, \rho_0), A_f \in G_{\rho_0}\}$ as per Lagrangian dual method and gradient descent (see (\ref{fortable1}) and (\ref{fortable2}) for $\varrho=A_f$)\;
    
    \If{${A_f}^\ast$ $\in$ $G_{\rho_0}$}{
        Update $A_f$ to ${A_f}^\ast$\;
    }
    \Else{
        Stop\;
    }
    Increment $i$ by 1\;
}
\Repeat{convergence}{
    Fix $A_f$ at ${A_f}^\ast$ obtained from the previous step\;
    
    Calculate the optimal value of $\rho$ ($\rho^\ast$) for the convex problem by $\max\{F(A_f, \rho), \rho \in G_{{A_f}^\ast}\}$ as per Lagrangian dual method and gradient descent (see (\ref{fortable1}) and (\ref{fortable2}) for $\varrho=\rho$)\;
    
    \If{$\rho^\ast$ $\in$ $G_{{A_f}^\ast}$}{
        Update $\rho$ to $\rho^\ast$\;
    }
    \Else{
        Stop\;
    }
    Increment $i$ by 1\;
}
Calculate the maximum $R_s$ using ${A_f}^\ast$ and $\rho^\ast$ obtained in the previous steps\;

Output the maximum $R_s$, ${A_f}^\ast$, and $\rho^\ast$\
\end{algorithm}

 It is noteworthy to emphasize that the optimization process terminates when the gradient descent approach attains a state whereby the objective function exhibits little change, signifying convergence.  As mentioned in    Algorithm 1,   Lagrangian approach can be utilized to find the optimal value of $\rho$ and $A_f$. The Lagrangian of $\mathcal{P}1$ can be expressed as
\begin{IEEEeqnarray}{lCr} \label{fortable1}
 \mathcal{L} \left( \varrho, \Upsilon_1, \Upsilon_2, \Upsilon_3 \right)= R_s + \Upsilon_1 (\varrho-1)+  \Upsilon_2    (-\varrho)+\nonumber \\\Upsilon_3(R_{pt}-R_p),
 \end{IEEEeqnarray}
 \noindent  where  $\Upsilon_1, \Upsilon_2,\; \text{and} \; \Upsilon_3 $ are the dual variables associated with the constraint on $\varrho$, for $\varrho \in \left(\rho, A_f\right)$, and the rate of PUs in (\ref{firstrate}), respectively.  Then, the Lagrange dual function of $\mathcal{P}1$ is given by
\begin{IEEEeqnarray}{lCr}  \label{fortable2}
 \mathcal{L} \left(  \Upsilon_1, \Upsilon_2, \Upsilon_3 \right)= \underset{ \varrho}{\text{max}}\;  \mathcal{L} \left(\varrho; \Upsilon_1, \Upsilon_2, \Upsilon_3 \right).
 \end{IEEEeqnarray}
\noindent Through the partial derivative and   the   the gradient descent method, the values of $\rho^{\ast}$, $A_f^{\ast}$,  $\Upsilon_1$, $\Upsilon_2$, and $\Upsilon_3$ are evaluated. 

\section{Numerical results}
The results of our theoretical investigations and Monte-Carlo simulations are detailed in this section. In this scenario, we assume that the relays are distributed out in an HPPP fashion throughout a 2D region $(U=2)$. In a 20-by-20-meter ($m$) square area, $10^5$ possible relay positions are generated. Throughout the results, we assume that $\kappa_{jp}=\kappa_{js}=\kappa$,  $\mu_{jp}=\mu_{js}=\mu$, and $T=1$.
 \begin{figure}  [t] 
  \centering
  \includegraphics[width=0.8\linewidth]{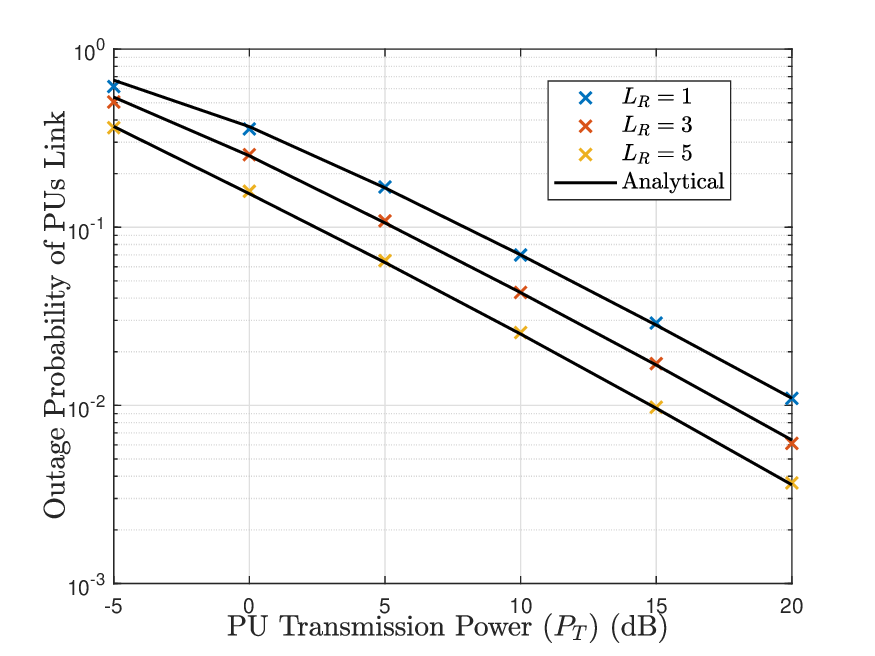}
  \caption{Outage probability of the PUs' link versus PU transmission power $(P_T)$.  $\rho=0.6$, $\delta=1$, $\alpha=2$, $\lambda_{PR}=0.5$, $n_p=n_s=2$, $\kappa=1$, $\mu=1$, $k=1$, $R_{thp}=0.5$ bits/sec/Hz, $L_S=2$,  $\eta=0.8$, $A_f=0.8$, $\phi=1$, $\nu_s=0$, and $\nu_p=0$.}
  \label{one}
\end{figure}

\begin{figure}  [t] 
  \centering
  \includegraphics[width=0.8\linewidth]{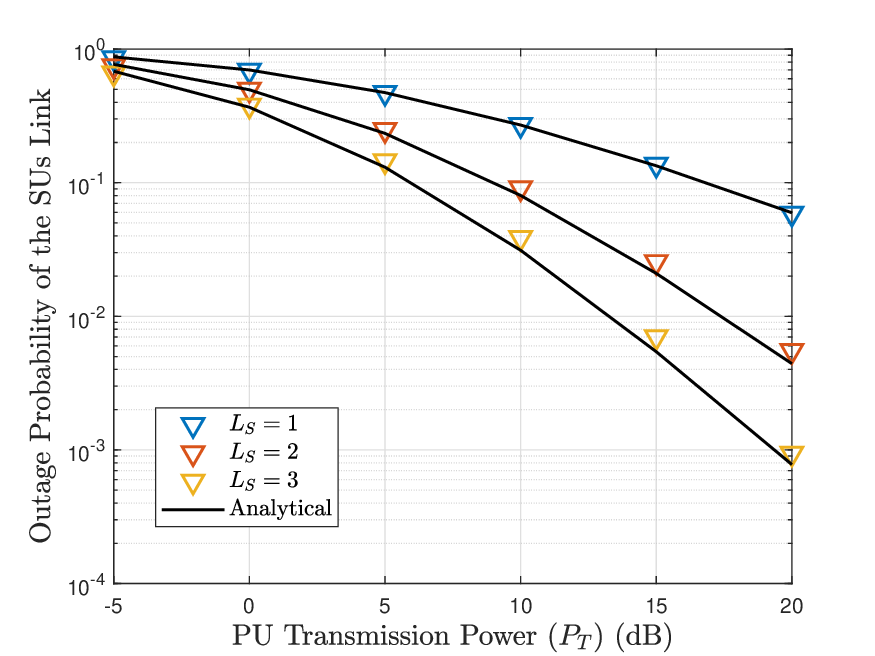}
  \caption{ Outage probability of the SUs' link versus PU transmission power $(P_T)$.  $\rho=0.2$, $\delta=1$, $\alpha=2$, $\lambda_{PR}=0.5$, $n_p=n_s=2$, $\kappa=1$, $\mu=1$, $k=1$, $R_{ths}=1$ bits/sec/Hz,   $\eta=0.8$, $A_f=0.2$, $\phi=1$, $\nu_p=0.2$, $\nu_s=0.2$, and $L_R=2$. }
  \label{two}
\end{figure}

 Referring to  Figure \ref{one}, for various values of the relaying SU's antennas $(L_R)$, the outage probability is shown as a function of the PU transmission power $(P_T)$. As the transmission power of a PU transmitter increases, the connection reliability of PUs seems to improve. This is because the higher the transmitted power, the greater the amount of energy gathered at the relay. Hence,  the higher the quality of the received messages at the PU receiver in terms of the received SINR. In addition, when the forwarding SU relay is equipped with a greater number of antennas, the communication between the PUs is enhanced. This is because the relay employs MRC technology, and the effectiveness of this method rises when more antennas are deployed.   Figure \ref{two} demonstrates that the PU transmission power may also improve the SUs' network performance.  This is because the amount of energy used specifically to transmit the SUs' messages increases as the SU relay gathers more energy. Moreover, as the SU receiver employs the MRC approach, the transmission reliability of SUs rises as the number of antennas $(L_S)$ increases. Additionally, when considering a fixed high value of $P_T$, the increase in system reliability caused by raising $L_S$ (the gap) is greater than the improvement for a fixed low value of $P_T$.  This demonstrates that the MRC method becomes more effective when the PU transmitter broadcasts messages at greater power levels.

\begin{figure}  [b] 
  \centering
  \includegraphics[width=0.8\linewidth]{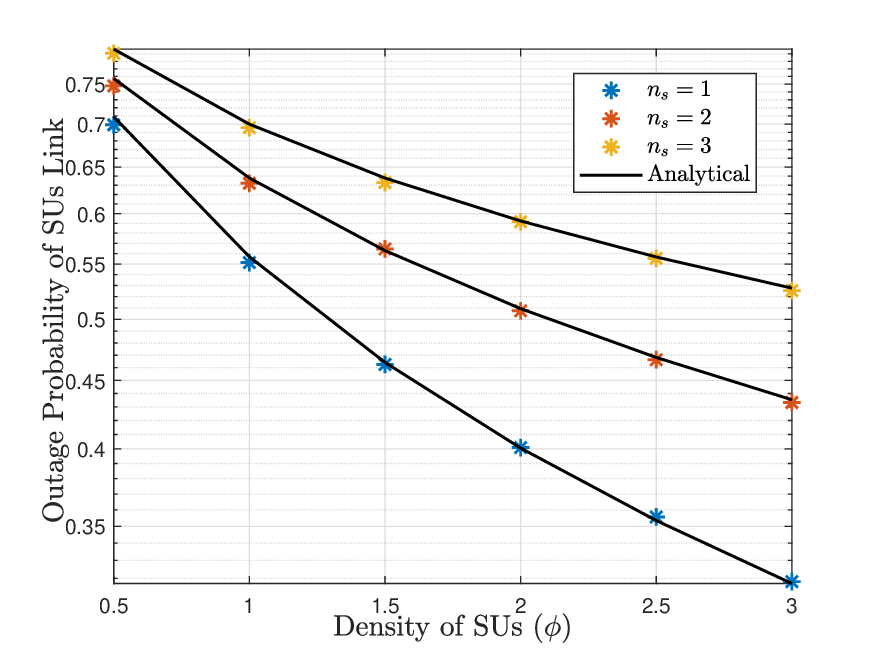}
  \caption{ Outage probability of the SUs' link versus SUs' density $(\phi)$.  $\rho=0.2$, $\delta=1$, $\alpha=2$, $\lambda_{PR}=0.5$,   $\kappa=0$, $\mu=1$, $k=1$, $R_{ths}=1$ bits/sec/Hz,   $\eta=0.8$, $A_f=0.2$,   $\nu_p=0.2$, $\nu_s=0.2$, $L_R=2$, $L_S=1$, $P_T=2$dB, and $n_p=1$.}
  \label{three}
\end{figure}

 Figure \ref{three} displays the impact of the density of the SU transmitters $(\phi)$ and the cascade level of the SUs connection $(n_s)$ on the performance of the SUs' transmission.  It is observed that  when more relays exist,  the possibility for the presence of an SU relay with reliable channel conditions for forwarding SUs' messages increases. Therefore, improving  the reception's quality. In addition, it can be seen that the cascade level $(n_s)$ has an influence that cannot be underestimated while assessing the system's performance. Particularly, when $n_s$ grows, meaning a higher number of scatters and objects present, the fading is more severe. This results in a poor  quality of the final received signal at the SU-Rx, hence, a higher outage probability.

\begin{figure}  [t] 
  \centering
  \includegraphics[width=0.8\linewidth]{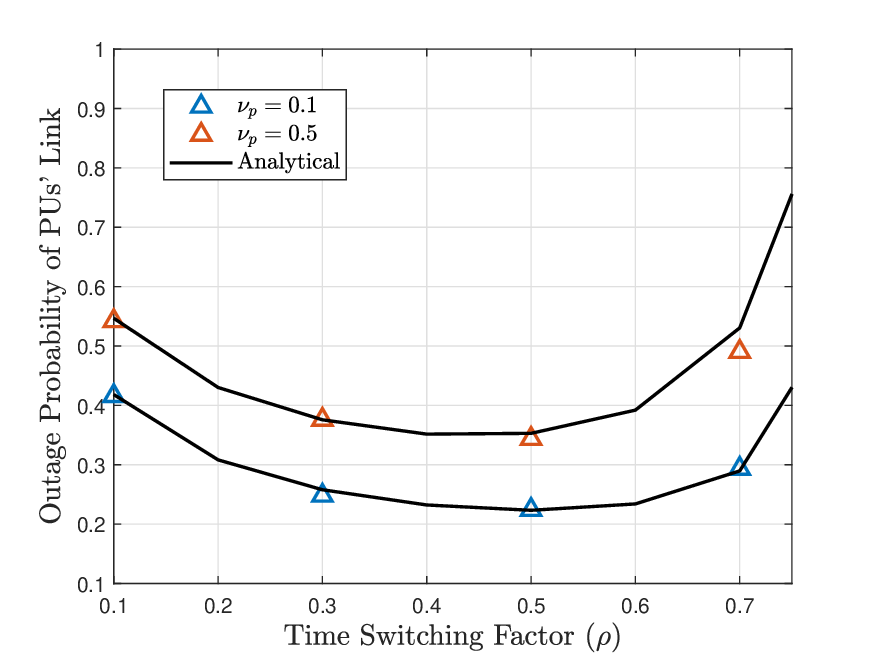}
  \caption{ Outage probability of the PUs' link versus the time switching factor $(\rho)$.    $\delta=1$, $\alpha=2$, $\lambda_{PR}=0.5$, $n_p=n_s=2$, $\kappa=1$, $\mu=1$, $k=1$, $R_{ths}=0.5$ bits/sec/Hz,   $\eta=0.8$, $A_f=0.9$,       $L_R=2$, $L_S=1$, $P_T=5$ dB, and $\phi=0.5$.}
  \label{three_2}
\end{figure}

The outage probability is a convex function of the time switching factor $(\rho)$, as illustrated by Figure \ref{three_2}. Particularly, when $\rho$ grows, reflecting that the time slot available for EH is increasing, the SU relay gathers more energy, resulting in more reliable transmission. After the minimal value of the outage probability is achieved, the system performance tends to degrade as the remaining time slot allocated for amplifying and forwarding the PUs' messages reduces. It is also observed that the reliability of the transmission decreases as the PS factor at the PU receiver $(\nu_p)$ rises. However, the storage device at PU-Rx contains energy that may be employed for subsequent transmissions, hence enhancing the energy efficiency of this model. This diagram illustrates that the receiver must find a balance between system reliability and energy content in its storage devices.

\begin{figure}  [b] 
  \centering
  \includegraphics[width=0.8\linewidth]{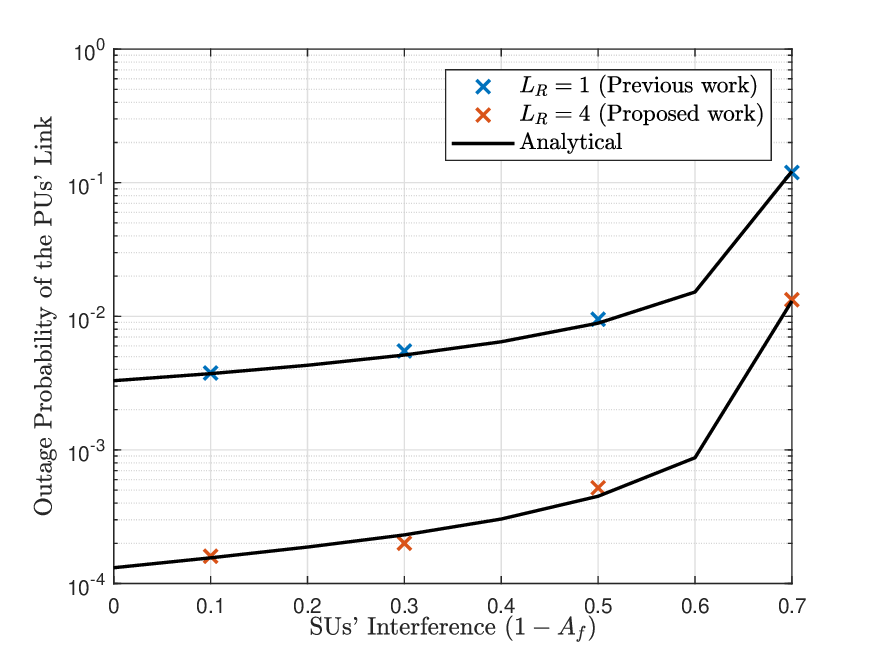}
  \caption{ Outage probability of the PUs' link versus the SUs' interference $(1-A_f)$.    $\delta=100$, $\alpha=2$, $\lambda_{PR}=1$, $n_p=n_s=1$, $\kappa=0$, $\mu=1$, $\nu_p=0$, $k=1$, $R_{thp}=0.2$ bits/sec/Hz,   $\eta=0.7$,         $P_T=5$ dB, and $\phi=100$.}
  \label{three_3}
\end{figure}

 The impact of SUs transmissions on PUs communication is demonstrated in  Figure \ref{three_3}. It is observed that as the SU relay allocates more power to convey the SUs' messages in terms of $1-A_f$, the outage probability at the PUs' link rises. This is attributed to the fact that  the PU receiver interprets the SUs' transmissions as interference. Furthermore, setting $n_p=1$, $L_R=1$,  $\kappa=0$, and $\mu=1$ (Rayleigh fading model) reduces this work to the one in \cite{9926102}, demonstrating the generality and accuracy of our suggested work. Additionally, when $L_R=4$ is being used, it is clear that our approach is superior to the one reported in \cite{9926102}.

\begin{figure}  [t] 
  \centering
  \includegraphics[width=0.8\linewidth]{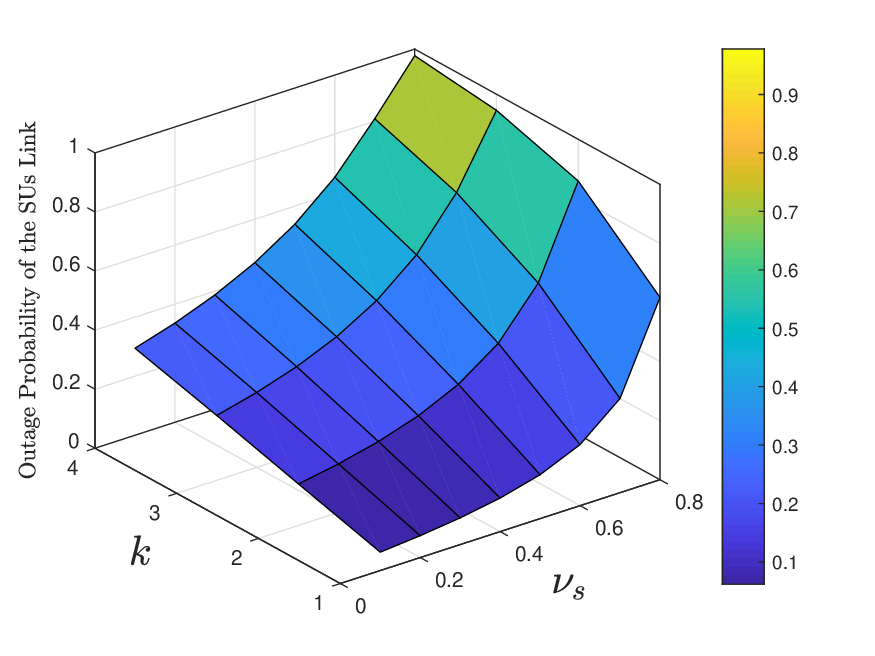}
  \caption{ Outage probability of the SUs' link versus   $k$ and $\nu_s$.  $\rho=0.2$, $\delta=1$, $\alpha=2$, $\lambda_{PR}=0.5$,  $n_p=n_s=2$, $\kappa=1$, $\mu=1$,   $R_{ths}=1$ bits/sec/Hz,   $\eta=0.8$, $A_f=0.1$,    $\nu_p=0.2$, $L_R=2$, $L_S=3$, $P_T=5$ dB, and  $\phi=1$.}
  \label{four}
\end{figure}

Figure \ref{four} illustrates a three-dimensional (3D) representation of the outage probability versus the $k^{th}$ nearest SU relay to the PU transmitter $(k)$ and the power splitting factor at the SU receiver $(\nu_s)$. As the SU receiver allocates more energy for EH, i.e., as $\nu_s$ grows, the risk of an outage rises since less energy is left for information decoding $(1-\nu_s)$. In this context, the receiver may determine the value of $\nu_s$ based on its needs, such as boosting system reliability or raising its energy level in order to exploit the stored energy for following transmissions.
As $k$ rises, indicating the selection of a distant SU relay, the system performance declines. This is due to the fact that when adopting a remote SU (e.g., $k=4$), the channel becomes less reliable in comparison to a nearby SU, i.e., for $k<4$. Notably, although selecting the nearest SU would be the optimal choice, assuming multiple SUs still provides a significant benefit. In other words, the fourth nearest SU may be assumed ($k=4$) if the first, second, and third closest SUs have unstable channels with the PU transmitter. This implies that several SUs ensure the continuation of communication between the two networks. Importantly, it is seen in the figure that if only the fourth closest one is available, the SU may collect less energy, i.e., reduce $\nu_s$, in order to maintain the transmission performance.

    \begin{figure}  [b] 
  \centering
  \includegraphics[width=0.8\linewidth]{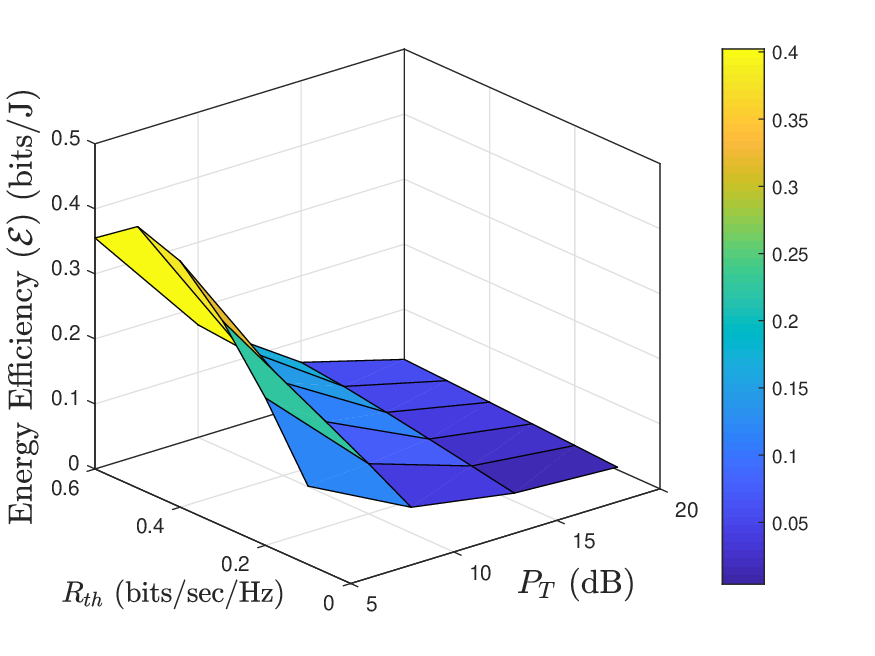}
  \caption{ Energy Efficiency   versus   $P_T$ and $R_{th}$.  $\rho=0.2$, $\delta=1$, $\alpha=2$, $\lambda_{PR}=0.5$,  $n_p=n_s=1$, $\kappa=0$, $\mu=1$,   $R_{ths}=R_{thp}=R_{th}$, $k=1$,  $\eta=0.8$, $A_f=0.5$,    $\nu_p=0.1$, $\nu_s=0.1$, $L_R=2$, and $L_S=1$,     $\phi=1$.}
  \label{five}
\end{figure}
 
 Figure \ref{five} depicts a 3D representation of the relationship between the energy efficiency $(\mathcal{E})$ in bits/Joule (J), the PU transmission power $(P_T)$, and the target rate $(R_{th})$. It is evident that when the target rate is high, the system achieves greater energy efficiency for low to moderate $P_T$ values. In light of this, it may be stated that maximum energy efficiency is only obtained at specified values of $P_T$, which depend on the required target rate. Particularly, $P_T$ values should shift towards   lower values as the target rate rises in order for the system to attain its maximum energy efficiency.


       \begin{figure}  [t] 
  \centering
  \includegraphics[width=0.8\linewidth]{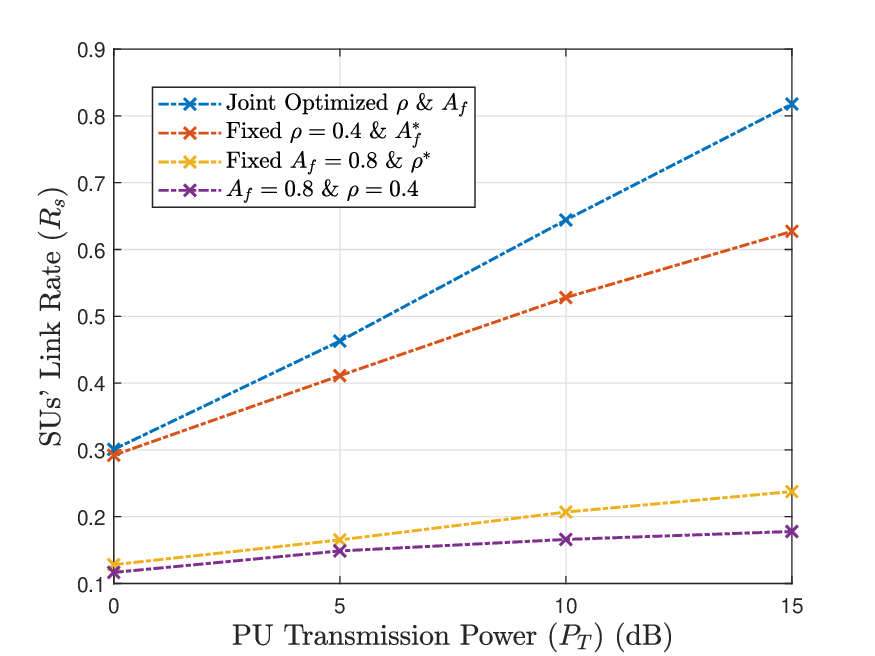}
  \caption{  SUs' link rate versus  PU transmission power $(P_T)$.  $\delta=1$, $\alpha=2$, $\lambda_{PR}=0.5$,  $n_p=n_s=2$, $\kappa=0$, $\mu=1$,   $R_{ths}=R_{thp}=R_{th}$, $k=1$,  $\eta=0.8$,     $\nu_p=0$, $\nu_s=0$, $L_R=2$, $L_S=1$, $R_{pt}=0.4$, and   $\phi=1$.  }
  \label{six}
\end{figure}

 From Figures \ref{three_2} and \ref{three_3}, it is evident that optimizing the parameters $\rho$ and $A_f$ would be more advantageous for achieving optimal system reliability. This is because $\rho$ determines the amount of time devoted to EH and, therefore, the amount of time available for amplifying and transmitting signals. In addition, optimizing $A_f$ would aid in determining how much power should be provided to forward each network's messages without producing intolerable damage to the other network. Consequently, Figure \ref{six} demonstrates that jointly optimizing these two parameters considerably increases the rate of SUs while maintaining the rate of PUs above a predetermined threshold $(R_{pt})$. In addition, compared to a single optimization for either $\rho$ or $A_f$ and their fixed values, the performance of our suggested optimization problem is superior.


\section{Conclusions}
This article investigates an overlay CRN with two PUs and multiple SUs.  These  SUs are spread randomly, and one is chosen to act as a relay for the PUs. In return, SUs gain access to the licensed band to forward their communications. The relay and both receivers collect energy using the TS protocol and the PS protocol, respectively. The receiving links are modeled using cascaded $\kappa$-$\mu$ channels on the assumption that they would experience several scatters. Due to the use of the MRC approach, the utilization of multiple antennas enhances the network's reliability, according to the results. By increasing the number of dispersed SU relays, the reliability of the PUs and SUs may also be strengthened.  In addition, the results show that the cascaded channels are more practical to presume in wireless channels. This is because they have a major influence on the reliability of the system that cannot be ignored, particularly for vehicular systems. In addition, while adopting EH by PU-Rx and SU-Rx decreases transmission reliability, it ensures the availability of energy in the devices' storage components for subsequent transmissions. To ensure optimal energy efficiency, the transmission power must be carefully chosen based on the target data rates.
In addition, the rate of SUs was maximized by jointly optimizing the time switching and power allocation factors at the relay, while maintaining an acceptable data rate for PUs. Despite our protocols' solid basis, further work integrating machine learning (ML) is anticipated to improve performance and mitigate any deterioration caused by the present approach's absence of ML.  As a possible future work, a reinforcement learning technique can be used to handle the dynamic environment of the considered cognitive vehicular network to enhance adaptability and intelligent decision-making.


\end{document}